# Analytical solution to the Langmuir spherical problem

**Dimitar G Stoyanov**
Sofia Technical University, Sliven Engineering and Pedagogical Faculty, 59 Burgasko Shosse Blvd, 8800 Sliven, Bulgaria

E-mail: dgstoyanov@abv.bg

**Abstract**
The current paper discusses and analytically solves the Langmuir spherical problem. A general solution has been obtained in a parametric representation and expressed in terms of the Airy function. A solution to the electric potential in a spherical vacuum diode with limited electron flow has also been reached.

## 1. Introduction

The present state of knowledge and developed technologies in control and application of charged particle beams has marked its beginning since Child and Langmuir pioneer efforts [1-4]. In their works, they were the first to present problem and the approach as well as solutions for different space configurations. Later, their ideas were adopted and applied to the production of vacuum lamps, TV kinescopes, etc. Nowadays, in the age of microelectronics, these components have become obsolete. However, the principles used then seem to be contemporary in charged particle accelerators, technological electronic guns, ionic engines, in spaceships for interplanetary flights.
   To improve such devices with charged particle beams or to develop new ones necessitates for the control theory of charged particle beams to be extended. On the one hand, for the last fifty years, remarkable success has been achieved in research and development of collective processes of ensembles of particles. Thus, gained experience makes it possible to look upon old problems in a different way and to consider them from a new perspective. On the other hand, the researchers' computing power has increased due to computers. Nowadays, researchers can modeling different phenomena easily and can apply any mathematical operations, special functions, etc. Therefore, old problems must be reconsidered and mathematically modernized to find analytical dependences, which are time and resource saving. Hence, models have become complicated, depicting the exact and correct alignments of Nature.
The present paper aims at finding an analytical solution to the Langmuir spherical problem for mono-energetic electrons [4] with initial velocity [5]. The Langmuir spherical problem [4] is expressed by a differential equation. However, an analytical solution has not been given to this differential equation, which is second-order and nonlinear. Noteworthy is that Langmuir had presented the solution in a table characterized by high accuracy for that time.

## 2. Description of the problem
### *2.1 Expressing the equation*
The spherical vacuum diode is a system of two concentric spherical metal electrodes. The inner electrode, with a radius of $r_k$, acts as a cathode of the diode. Outer electrode with a radius of $r_a$ acts as an anode ($r_a > r_k$). There is a high vacuum in the volume between the two spheres (absolute permittivity of vacuum is $\varepsilon_0$).

The cathode emits electrons with a mass, **m**, an electrical charge, **q** ($q < 0$), and initial velocity, $V_0$, outwarded radially. The current density of the emitted electrons on the cathode surface is equal. Thus, the emitted electrons (they movement) creates current, $I_0$. Both metal electrodes are below the potential difference of $U = \varphi_a - \varphi_k$. Here $\varphi_k$ represents the cathode electric potential while $\varphi_a$ stands for the anode potential.

The electron movement and the parameters of the electric field will be described using a spherical coordinate system with an origin lying at the centre of the spheres $\vec{r} = (r, \beta, \theta)$. Due to the spherical symmetry of the problem, all scalars depend only on the coordinate **r**. All vectors of the values are radial and depend on the coordinate **r**.

The electrical potential $\varphi(r)$ in the volume closed by the two spherical electrodes as a function of the potential is as follows

$$\varphi(r = r_k) = \varphi_k \tag{1}$$

$$\varphi(r = r_a) = \varphi_a \tag{2}$$

The electrical field potential is formed by the volume density of the electron charge of **k** and is expressed by the Poisson's equation. Due to spherical symmetry the latter takes the form [4]

$$\frac{1}{r^2} \cdot \frac{d}{dr}\left(r^2 \frac{d\varphi}{dr}\right) = -\frac{k}{\varepsilon_0} \tag{3}$$

The emitted cathode electrons flow into the volume of the two electrodes and move at a velocity of $V_r$. Due to the conservative electric field, which acts on the electron, the law of conservation of energy is valid for the electron. To calculate the velocity value the law of conservation of energy will be applied for the cathode surface as well as for an arbitrary point of the volume at the distance of **r** from the origin of the coordinate system.

$$\frac{m.V_r^2}{2} + q.\varphi(r) = \frac{m.V_0^2}{2} + q.\varphi_k \tag{4}$$

From (4) we obtain [5]

$$V_r = \sqrt{V_0^2 + \frac{2.q}{m} \cdot (\varphi_k - \varphi(r))} \tag{5}$$

The charged particles move and create current with a density of $j_r$.
For the current density we have

$$j_r = k.V_r \tag{6}$$

From the law of conservation of charge for the stationary case we obtain

$$\frac{1}{r^2} \cdot \frac{d}{dr}\left(r^2 . j_r\right) = 0 \tag{7}$$

In respect with (7) the product in the derivative is a constant quantity. Thus, we have

$$j_r = -\frac{I_0}{4.\pi.r^2} \tag{8}$$

Equation (6) regarding (5) and (8) gives

$$k = \frac{j_r}{V_r} = -\frac{I_0}{4.\pi.r^2.\sqrt{V_0^2 + \frac{2.q}{m}.(\varphi_k - \varphi(r))}} \quad (9)$$

So far, the volume density of the electrical charge, $k$, has been expressed by the current flowing along the diode, $I_0$, and the electron velocity, $V_r$. Replacing (9) in (3) we obtain

$$r^2 \frac{d^2\varphi}{dr^2} + 2r\frac{d\varphi}{dr} = +\frac{I_0}{4.\pi.\varepsilon_0.\sqrt{V_0^2 + \frac{2.q}{m}.(\varphi_k - \varphi(r))}} \quad (10)$$

The result is a second-order ordinary differential equation for the electric field potential $\varphi(r)$ in the volume of a spherical vacuum diode. Considering the fact the equation is nonlinear and its solution involves difficulties.

**2.2** *Making the equation dimensionless*
The first step in finding a solution to the equation is to make it dimensionless. The initial electron kinetic energy $E_0$ [5] will be taken as a scale of energy

$$E_0 = \frac{m.V_0^2}{2} \quad (11)$$

Defining the dimensionless potential we have [5]

$$\Phi = \frac{q.(\varphi_k - \varphi(r))}{E_0} \quad (12)$$

Replacing (12) in (10) we obtain

$$r^2 \frac{d^2\Phi}{dr^2} + 2r\frac{d\Phi}{dr} = \frac{-q}{V_0.E_0} \frac{I_0}{4.\pi.\varepsilon_0.\sqrt{1+\Phi}} \quad (13)$$

The constant coefficients on the right side of the equation (13) can be combined in one and be considered as dimensionless current, $i_0$

$$i_0 = \frac{-q}{V_0.E_0} \frac{I_0}{4.\pi.\varepsilon_0} = \frac{-2q}{m}.\frac{1}{V_0^3}.\frac{I_0}{4.\pi.\varepsilon_0} \quad (14)$$

Then (13) can be rewritten as

$$r^2 \frac{d^2\Phi}{dr^2} + 2r\frac{d\Phi}{dr} = \frac{i_0}{\sqrt{1+\Phi}} \quad (15)$$

Henceforth, we shall solve this equation.

## 3. Solution to the equation
### 3.1 Langmuir solution

Langmuir [4] used the logarithm from the relation between the radius and the emitter (cathode) radius as a space parameter. Assuming the electrical field intensity on the emitter is zero this radius corresponds to the radius of the minimum electric potential $r_m$. Respectively, we have

$$\eta = \ln\left(\frac{r}{r_m}\right) \quad (16)$$

Applying (16) the equation (15) can be rewritten as

$$\sqrt{1+\Phi} \cdot \left(\frac{d^2\Phi}{d\eta^2} + \frac{d\Phi}{d\eta}\right) = i_0 \quad (17)$$

The function of the dimensionless potential $\Phi(\eta)$ is expressed by the dimensionless function $\alpha(\eta)$ as follows [4]:

$$(1+\Phi)^{3/2} = \frac{9}{4} \cdot i_0 \cdot \alpha^2 \quad (18)$$

Then, the potential is expressed as

$$(1+\Phi) = \left(\frac{9}{4} \cdot i_0\right)^{2/3} \cdot \alpha^{4/3} \quad (19)$$

Substituting it in (17) and calculating it we have

$$3.\alpha.\alpha'' + \alpha'^2 + 3.\alpha.\alpha' = 1 \quad (20)$$

Here prime represents a derivative with respect to the function argument.

**Table 1.** Values of $\alpha^2(r)$ - solution to (20) [4].

| $r/r_m$ | $\alpha^2(r)$ | $r_m/r$ | $\alpha^2(r)$ |
|---------|---------------|---------|---------------|
| 1.0 | 0.0000 | 1.0 | 0.0000 |
| 1.1 | 0.0086 | 1.1 | 0.0096 |
| 1.2 | 0.0299 | 1.2 | 0.0372 |
| 1.5 | 0.1302 | 1.5 | 0.2118 |
| 2.0 | 0.3260 | 2.0 | 0.7500 |
| 5.0 | 1.1410 | 5.0 | 7.9760 |
| 10.0 | 1.7770 | 10.0 | 29.1900 |
| 20.0 | 2.3780 | 20.0 | 93.2400 |
| 50.0 | 3.1200 | 50.0 | 395.3000 |
| 100.0 | 3.6520 | 100.0 | 1 141.0000 |
| 200.0 | 4.1660 | 200.0 | 3 270.0000 |
| 500.0 | 4.8290 | 500.0 | 13 015.0000 |

Thereby, we obtain the differential equation for $\alpha(\eta)$. It is a second-order nonlinear differential equation. Yet nobody has ever found an analytical solution to it. Therefore, it will be represented in the form of Taylor infinite series. The first terms of the series give [4]

$$\alpha^2 = \eta^2 - 0.6\eta^3 + 0.24\eta^4 - 0.074\eta^5 + ..... \tag{21}$$

Langmuir solved the equation (20) numerically and presented $\alpha(\eta)$ in a table. Some of the values [4] have been given in Table 1 as an example.

### 3.2 New general solution
We start by transforming the left side of the equation (15) into an equivalent form.
We obtain:

$$r\frac{d^2}{dr^2}(r.(1+\Phi)) = \frac{i_0}{\sqrt{1+\Phi}} \tag{22}$$

Substituting

$$(1+\Phi)^{3/2} = 2.i_0.f^{3/2} \tag{23}$$

$$u = \frac{d}{dr}[r.f] \tag{24}$$

in (22) we obtain

$$r\frac{du}{dr} = \frac{1}{2\sqrt{f}} \tag{25}$$

From where

$$\frac{d\sqrt{r}}{du} = \sqrt{r.f} \tag{26}$$

We differentiate (26) with respect to **u** and obtain

$$\frac{d^2\sqrt{r}}{du^2} = \frac{d\sqrt{rf}}{du} = \frac{1}{2.\sqrt{r.f}} \cdot \frac{d[r.f]}{dr} \cdot \frac{dr}{du} = \sqrt{r}.u \tag{27}$$

Finally we have

$$\frac{d^2\sqrt{r}}{du^2} = \sqrt{r}.u \tag{28}$$

The general solution of the equation is expressed [6] by the Airy functions

$$\sqrt{r} = C_1.Ai(u) + C_2.Bi(u) \tag{29}$$

And (26) leads to

$$\sqrt{rf} = \frac{d\sqrt{r}}{du} = C_1.Ai'(u) + C_2.Bi'(u) \tag{30}$$

Here prime represents the derivative of the function with respect to the argument.
This is a parametric representation (using the parameter $u$) of the solution of (22).
Here $C_1$ and $C_2$ are constants of integration, determined by the initial and boundary conditions.

*3.3 Limited electron flow*
First, (24) will be considered carefully. The development of the derivative gives us

$$u = f + r\frac{df}{dr} \tag{31}$$

Considering the potential minimum of $f_m$ at $r = r_m$, the potential derivative with respect to $r$ in this point is zero. Hence, we abtain

$$u_m = f_m \tag{32}$$

The value of $f_m$ is a non-negative number, which is positive for unlimited electron flow, whereas it has a zero value in limited current [5].
We need to examine the solutions for which $f_m$ and $u_m$, according to (32), have a zero value. Thus, the exact values of the integration constants will be expressed.
Provided $f_m$ is zero and $u_m$ is zero, from (30) we obtain

$$C_1.Ai'(0) + C_2.Bi'(0) = 0 \tag{33}$$

Therefore, we have

$$\frac{C_1}{Bi'(0)} = -\frac{C_2}{Ai'(0)} = C_3 \tag{34}$$

Replacing (34) in (29) and (30) we obtain

$$\sqrt{r} = C_3.[Bi'(0).Ai(u) - Ai'(0).Bi(u)] \tag{35}$$
$$\sqrt{rf} = C_3.[Bi'(0).Ai'(u) - Ai'(0).Bi'(u)] \tag{36}$$

If we replace the parameter of $u$ with zero in (35), we obtain

$$\sqrt{r_m} = C_3.[Bi'(0).Ai(0) - Ai'(0).Bi(0)] \tag{37}$$

Thus, we can obtain the value of the constant $C_3$ and get to

$$\sqrt{\frac{r}{r_m}} = \frac{Bi'(0).Ai(u) - Ai'(0).Bi(u)}{Bi'(0).Ai(0) - Ai'(0).Bi(0)} \tag{38}$$

$$\sqrt{f} = \frac{Bi'(0).Ai'(u) - Ai'(0).Bi'(u)}{Bi'(0).Ai(u) - Ai'(0).Bi(u)} \tag{39}$$

The range of variations for $\mathbf{u}$ is determined by the positive number of the right side of (38). Therefore, $\mathbf{u} \in (-1.9864, +\infty]$.

## 4. Consideration of solution

In this part, we are going to introduce a new quantity of $\mathbf{G(r)}$ to make considerations simple and to compare the two functional dependences of the potential (18) and (23). They take the form

$$G_1(r) = \frac{(1+\Phi)^{3/2}}{i_0} = \frac{9}{4}\cdot\alpha^2 \tag{40}$$

$$G_2(r) = \frac{(1+\Phi)^{3/2}}{i_0} = 2\cdot f^{3/2} \tag{41}$$

The two dependences are shown in figure 1. The curve represents (41), whereas each circle shows (40) for some of the data from Table 1 (for values of $\alpha^2(\mathbf{r})$).

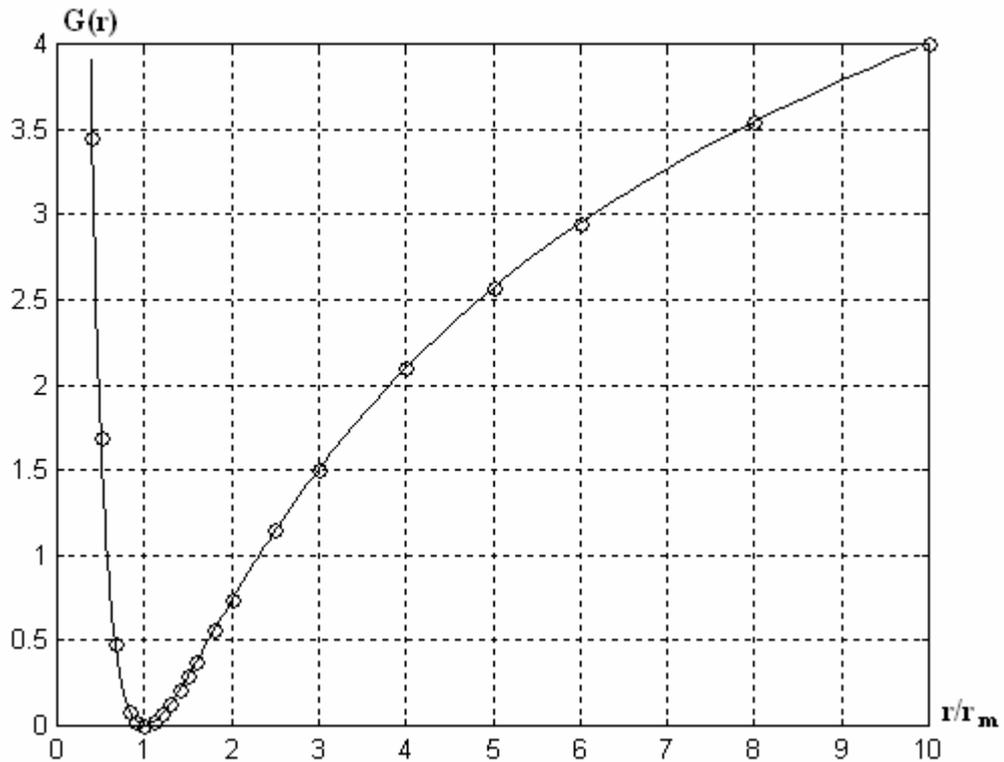

**Figure 1.** A graph of the $G_2(r)$ function after (41) (the continuous curve) and the values for $G_1(r)$ after (40) and Table 1 (the circles).

The graph displays the correspondence between the analytic al solution obtained here and Langmuir tabular values.

Furthermore, a new special method for defining $\alpha^2(\mathbf{r})$ can be given.

First, we define the auxiliary function of $\mathbf{F(u)}$ as the parameter of $\mathbf{u}$ is not limited

$$F(u) = \frac{Bi'(0)\cdot Ai(u) - Ai'(0)\cdot Bi(u)}{Bi'(0)\cdot Ai(0) - Ai'(0)\cdot Bi(0)} \tag{42}$$

Then

$$\frac{r}{r_m} = F^2(u) \qquad (43)$$

$$\alpha^2(r) = \frac{8}{9} \cdot \left(\frac{F'(u)}{F(u)}\right)^3 \qquad (44)$$

Here prime represents the derivative at the argument.

This is a parametric representation of $\alpha^2(r)$ (using the parameter $u$) by the function of (42). The range of variations of $u$ is determined by (42) which is a positive number. Therefore, $u \in (-1.9864, +\infty]$.

## 5. Conclusion

To sum up, the paper discusses Langmuir spherical problem. The differential equation for the electric potential has been obtained for monoenergetic electrons with a non-zero initial velocity. The obtained general analytical solution is a parametric solution which is expressed by the Airy function in limited electron flow. The particular solution for the electric potential in spherical vacuum diode has been reached. The analytical solution for the electric field potential in spherical vacuum diode has been compared to Langmuir tabular values. A major correspondence between them has been discovered.